\documentclass[british]{article}  
\usepackage{amsmath,amssymb,babel,mathrsfs,ae,aecompl,hyperref} 
\newcommand{\bv}{\vec v}
\newcommand{\bh}{\vec h}
\newcommand{\bk}{\vec k}

\title{Poincar\'e Covariance and \(\kappa\)-Minkowski Spacetime}
\author{Ludwik D\k{a}browski\thanks{dabrow@sissa.it}{}\ \  and Gherardo Piacitelli\thanks{gherardo@piacitelli.org}\\{\footnotesize SISSA -- Trieste, via Bonomea 265, 34136 Italia}} 

\begin{document} 
\maketitle
\begin{abstract}
A fully Poincar\'e covariant model is constructed as an extension of the 
\(\kappa\)-Minkowski spacetime. 
Covariance is implemented by a unitary representation of  the Poincar\'e group, 
and thus complies with the original Wigner approach to quantum  symmetries.
This provides yet another example (besides the DFR model),
where Poincar\'e covariance is realised \`a la Wigner in the presence 
of two characteristic dimensionful 
parameters: the light speed and the Planck length. 
In other words, a Doubly Special Relativity (DSR) framework  may well be realised 
without deforming the meaning of ``Poincar\'e covariance''.
\end{abstract}

\maketitle

\section{Introduction}
According to DSR scenario \cite{AmelinoCamelia:2000ge,AmelinoCamelia:2000mn}, 
to account for phenomena at Planck
scale (possibly in view of a theory of quantum gravity), ordinary Poincar\'e 
covariance should be properly modified, so to incorporate two universal
dimensionful parameters: the light speed \(c\sim 3\times 10^8\;
\text{cm}\cdot\text{s}^{-2}\) and the Planck length 
\(\lambda_P\sim 10^{-33}\;\)cm. 
It has been proposed \cite{KowalskiGlikman:2001gp} 
to test such a scenario on the so-called
\(\kappa\)-Minkowski spacetime, where the name refers to the notation 
\(\kappa=1/\lambda_P\) \cite{Lukierski:1991pn,Majid:1994cy}.
In that model, Poincar\'e covariance is deformed in the sense of quantum 
groups.  

While equations which are not form--invariant may well be compatible with
the principle of equivalence of Lorentz observers (e.g.\ the Coulomb gauge), 
a deformation of the notion of covariance  
at Planck scale would allow to select a privileged class of observers.

Moreover, the Doplicher-Fredenhagen-Roberts (DFR) model \cite{Doplicher:1994zv,Doplicher:1994tu}, proposed in 1994, shows that the usual transformation laws
of special relativity may well be ``doubly special'', without any deformation. 
In fact, therein 
the commutation relations with the corresponding quantisation 
of ordinary functions depend on \(\lambda_P\), and they are  
covariant under the adjoint  action of unitary operators representing 
the ordinary (undeformed) Poincar\'e group.
This contrasts the widespread folklore that the presence of 
two universal dimensionful parameters should necessarily 
force us into the realm of 
``modified'' Poincar\'e covariance. 

In this letter we provide another example
of a model with Poincar\'e covariance \`a la Wigner, and two dimensionful
parameters (\(c=\kappa=1\), in natural units). The (non trivial) 
defining relations are
\begin{subequations}
\label{eq:cov_rel}
\begin{gather}
\label{eq:cov_rel_1}
[X^\mu,X^\nu]=i\big(V^\mu(X-A)^\nu-V^\nu(X-A)^\mu\big),\\
\label{eq:cov_rel_4}
V_\mu V^\mu=I,
\end{gather}
\end{subequations}  
where \(A^\mu,V^\mu\) are central, and 
the usual Lorentz metric \(g^{\mu\nu}=\text{diag}(1,-1,-1,-1)\) is used.
Commutation relations 
are understood to hold strongly;
in particular the regular (Weyl)
form of \eqref{eq:cov_rel_1} is
\begin{equation}
\label{eq:new_weyl_rel}
e^{ihX}e^{ikX}=e^{i(h+k-\varphi(h,k;V))_\mu A^\mu}e^{i\varphi_\mu(V;h,k)X^\mu},
\end{equation}
where \(\varphi\) is given by equations \eqref{eq:psi}. 

More precisely, we claim
that it is possible to construct the selfadjoint operators \(X^\mu,V^\mu,A^\mu\)  together with
a unitary representation \(U\) of the full Poincar\'e group \(\mathscr P\),
such that
\begin{subequations}  
\label{eq:covariance}
\begin{gather}
U(\varLambda,a)^{-1}X^\mu U(\varLambda,a)={\varLambda^\mu}_\nu X^\nu+a^\mu I,\\
U(\varLambda,a)^{-1}V^\mu U(\varLambda,a)={\varLambda^\mu}_\nu V^\nu,\\
U(\varLambda,a)^{-1}A^\mu U(\varLambda,a)={\varLambda^\mu}_\nu A^\nu+a^\mu I
\end{gather}
\end{subequations}  

This is obtained by ``full covariantisation'' 
of the \(\kappa\)-Minkowski
model. The intermediate step of Lorentz covariantisation was discussed in 
\cite{Dabrowski:2009mw}, in a more general setting. The covariantised model contains the initial model 
as a component; in a sense (made precise in \cite{Dabrowski:2009mw}) 
it is the smallest possible
covariant central extension of the usual \(\kappa\)-Minkowski model. Note that, analogously, the DFR model could be equivalently
described as the minimal central full
covariantisation of a ``canonical quantum spacetime''. The procedure
is inspired
by the mathematical construction of crossed products (also known as 
covariance algebras). 

We take the point of view of the DFR analysis, where only the 
commutation relations among the coordinates (and their commutators) 
have to be modified; 
momenta (defined as the generators of translations) 
forcefully commute pairwise, while their 
commutation relations with the coordinates simply express the action
of infinitesimal translations of the coordinates, as usual. 
In other words, we quantise the 
configuration space, which amounts to 
find a candidate for replacing the 
localisation algebra of quantum field theory.
We do not quantise the phase space, namely we do not aim to
a ``more non-commutative'' version of quantum mechanics 
(see \cite{Doplicher:1994tu} for a 
general discussion about motivations, and \cite{Piacitelli:2010bn} 
for a review). Quantum field theoretical aspects will be discussed elsewhere.

\section{The Covariant Coordinates}
Covariant representations of the
relations \eqref{eq:cov_rel} will necessarily be highly reducible;
let us first  take for \(X^\mu,V^\mu,A^\mu\) 
an irreducible regular representation. By the Schur's lemma, 
\(V^\mu=v^\mu I,A^\mu=a^\mu I\) for some real 4-vectors \(v,a\), so that
\begin{equation}\label{eq:reduced_cov}
[X^\mu,X^\nu]=i(v^\mu (X^\nu-a^\nu I)-v^\nu (X^\mu-a^\mu I))
\end{equation}
and \(v_\mu v^\mu=1\). 
We recognise as a special case the usual \(\kappa\)-Minkowski
relations, corresponding to the choice \(v=v_{(0)},a=0\), where
\[
v_{(0)}=(1,0,0,0).
\]
For this particular case, the irreducible representations are all known 
\cite{Dabrowski:2009mw}: \(X_{(0)}^\mu\) will denote the corresponding universal
representation (containing all irreducibles precisely once), see the appendix. 
The operators \(X_{(0)}^\mu\) act on the Hilbert space \(\mathfrak H_{(0)}\), with scalar product \((\cdot,\cdot)_{(0)}\).

Now, for every \((\varLambda,a)\in\mathscr P\), the operators 
\[
X^\mu=(\varLambda X_{(0)}+aI)^\mu={\varLambda^\mu}_\nu X^\nu_{(0)}+a^\mu I
\] 
fulfill
\eqref{eq:reduced_cov} with \(v=\varLambda v_{(0)}\). In this way, it is possible
to obtain representation for any pair \((v,a)\in H\times\mathbb R^4\), where
\(H=\mathscr Lv_{(0)}\) is the orbit of \(v_{(0)}\) 
under the full Lorentz group 
\(\mathscr L=O(1,3)\); \(H\) is a two sheeted hyperboloid. 

By taking a direct integral over the Haar measure \(d(\varLambda,a)\)
of \(\mathscr P\)
of all the irreducible representations so constructed, it is easy
(see e.g.\ the analogous discussion of \cite{Doplicher:1994tu,Dabrowski:2009mw})
to construct selfadjoint operators \(X^\mu,V^\mu,A^\mu\) 
and a unitary representation \(U\)
of \(\mathscr P\), fulfilling (\ref{eq:cov_rel},\ref{eq:covariance}).
The result of the construction is equivalent to the following covariant
representation. Consider the Hilbert space of \(\mathfrak H_{(0)}\)-valued
functions \(\psi(\varLambda,a)\), with scalar product
\[
(\psi,\psi')=\int\limits_{\mathscr P} d(\varLambda,a)(\psi(\varLambda,a),\psi'(\varLambda,a))_{(0)}.
\]
Then set
\begin{subequations}
\begin {align}
(X^\mu\psi)(\varLambda,a)&=(\varLambda X_{(0)}+aI)^\mu\psi(\varLambda,a),\\
(V^\mu\psi)(\varLambda,a)&=(\varLambda v_{(0)})^\mu\psi(\varLambda,a),\\
A^\mu\psi(\varLambda,a)&=a^\mu\psi(\varLambda,a),\\
U(M,b)\psi(\varLambda,a)&=\psi((M,b)^{-1}(\varLambda,a)).
\end{align}
\end{subequations}
Above \((\varLambda,a)\) and \((M,b)\) are elements of \(\mathscr P\).
Note that \(U\) is a strongly continuous representation of \(\mathscr P\), and
we may define momentum operators \(P^\mu\) by setting
\(e^{ia^\mu P_\mu}=U(\mathbb I,a)\); they fulfil the commutation relations
\begin{subequations} 
\begin{gather}
\label{eq:abel_trans} 
[P^\mu,P^\nu]=0,\quad
[P^\mu,V^\nu]=0,\\
\label{eq:cov_trans} 
[P^\mu,A^\nu]=
[P^\mu,X^\nu]=ig^{\mu\nu}I.
\end{gather}
\end{subequations} 
Clearly, the first equation of \eqref{eq:abel_trans} follows from the 
abelianness of the translation subgroup; the second expresses translation invariance of \(V\), while
\eqref{eq:cov_trans} describes the effect of 
infinitesimal translations on \(X,A\). 

Analogously the generators of infinitesimal Lorentz transformations describe the usual (undeformed) action of the infinitesimal Lorentz transformations on
\(X^\mu,V^\mu,A^\mu,P^\mu\).  

Note also that the orthogonal projection 
\(E\psi=\chi\psi\), where
\(\chi\) is the characteristic function of the Lorentz subgroup \(\mathscr L\) of \(\mathscr P\),
fulfills \([E,X^\mu]=0\); it reduces the model to the Lorentz covariant 
component discussed in \cite{Dabrowski:2009mw}. Indeed, 
we have \([E,U((\varLambda,0))]=0\) always, even though \(E\) does 
not commute with \(U(\varLambda,a)\) for \(a\neq 0\). 

We remark that, at a formal level, the right hand side of 
\eqref{eq:reduced_cov} 
may be regarded as a combination of a ``canonical'' and
a ``Lie type'' contribution, according to a popular terminology.

\section{Weyl Symbols and Relations}
In order to define a star product, we consider functions \(f=f(v,a;x)\),
where \(x\) runs in the classical Minkowski spacetime \(\mathbb R^4\), while
\((v,a)\) runs in the parameter space \(H\times\mathbb R^4\) 
(surviving the classical limit as extra dimensions). Then the quantisation
prescription is
\[
f(V,A;X)=\frac1{(2\pi)^2}\int dk\;\hat f(V,A;k)e^{ikX}.
\]
Above, the replacement of the variables \(v^\mu,a^\mu\) by the operators
\(V^\mu,A^\mu\) respectively is understood in the usual sense of functions 
of pairwise commuting operators. The replacement of \(x\) by \(X\)
mimics instead the usual Weyl prescription, where
\[
\hat f(V,A;k)=\frac1{(2\pi)^2}\int dx\;f(V,A;x)e^{-ikx}.
\]

One may obtain the symbolic calculus with the star product defined by
\[
f(V,A;X)g(V,A;X)=(f\star g)(V,A;X).
\]
Quantisation intertwines operator adjunction and pointwise conjugation:
\[
\bar f(V,A;X)=f(V,A;X)^*.
\]

In order to carry over explicit computations, we need to know 
the Weyl relations \eqref{eq:new_weyl_rel} 
explicitly, namely to compute~\(\varphi\).
To this end, we first consider the irreducible case,
where \(V=vI,A=aI\). We use the standard space-time notations 
\(v=(v^\mu)=(v^0,\bv),\quad (v_\mu)=(v^0,-\bv)\), where
\(\bv\in\mathbb R^3\); the metric has signature \((+---)\).

We now think of \(\bv,\bh,\bk,\vec 0\) as row 3-vectors; 
\(\bv^t,\bh^t,\bk^t,\vec 0^t\) are the corresponding
column vectors (where \(t\) indicates row-by-column transposition). 
The Lorentz matrix
\[
\varLambda=\left(
\begin{array}{c|c}
v^0&\bv\\\hline\bv^t&\mathbb I+\frac{\bv^t\bv}{1+v^0}
\end{array}
\right)
\]
fulfils
\(\varLambda v^{(0)}=v\). We now want to compute the Weyl relations
for the operators \(X=\varLambda X_{(0)}+aI\).
We observe that
by definition, \(e^{ikX}=e^{ika}e^{i(\varLambda^{-1}k)X_{(0)}}\),
hence by (\ref{eq:old_kappa},\ref{eq:old_kappa_weyl}) we have
\begin{align*}
e^{ihX}e^{ikX}&=e^{i(h+k)a}e^{i\phi(\varLambda^{-1}h,\varLambda^{-1}k)X_{(0)}}\\
&=e^{i(h+k-\varphi(h,k;v))a}e^{i\varphi(h,k;v)X},
\end{align*}
where
\[
\varphi(h,k;v)=\varLambda\phi(\varLambda^{-1}h,\varLambda^{-1}k)).
\]

A routine computation yields
\begin{subequations}
\label{eq:psi}
\begin{align}
\nonumber
\varphi^0(h,k;v)=&
w((h+k)v, hv)e^{ikv}(v^0\vec h -h^0\vec v)\vec v^T+\\
\nonumber
&+ w((h+k)v, kv)(v^0\vec k -k^0\vec v)\vec v^T +\\
&+(h+k)v\, v^0 ,\\
\nonumber
\vec\varphi(h,k;v)=&
 w((h+k)v, hv)e^{ikv}(hv\, \vec v - \vec h)+\\
\nonumber
&+ w((h+k)v, kv)(kv\, \vec v - \vec kv)+\\
&- (h+k)v\, \vec v .
\end{align}
\end{subequations}
Note that, as expected, \(\varphi\) does not depend on the particular choice
of \(\varLambda\), provided that \(\varLambda v_{(0)}=v\).

To finally obtain \eqref{eq:new_weyl_rel} for the most general regular
representation,
it is sufficient to replace the variables \(v,a\) with the operators 
\(V,A\), in the usual sense of functions of pairwise commuting operators.

\section{Conclusions}
A universal length may well exist in a Poincar\'e covariant setting. There 
is no contradiction between this statement and the Lorentz-Fitzgerald 
contraction. Indeed, the Planck length
plays the r\^ole of a {\itshape characteristic} length ruling the structure
of commutation relations; it is not an observable quantity itself.
It follows that the quest for a universal length 
{\itshape alone} does not provide a motivation for deforming the 
notion of covariance.

Note that the approach to covariantisation followed here essentially consists
in extending the algebra of coordinates 
by a centre which provides the room where to accommodate the joint spectrum 
of the additional operators. This centre is already classical, and survives the large scale limit as a manifold of extra dimensions. See also \cite{MR1600364}.

It should be mentioned that this covariantisation method does not exploit
the symmetries which are possibly already present in the basic commutation 
relations. 
In particular, when applied to 
the \(\kappa\)-Minkowski, it does not extend covariance under the 
natural action of rotations on the initial set of generators. Covariance
is instead implemented  by an action of rotations (as elements 
of the Poincar\'e group) which replaces the original one.

By similar methods it is possible to covariantise 
the lightlike (\(V_\mu V^\mu=0\))
and spacelike (\(V_\mu V^\mu=-I\)) models proposed in \cite{Lukierski:2002ii}.
Indeed, this method is quite general and can be applied to a large class of 
models with Lie relations, provided that they have a good representation theory.

This however does not exhaust the possibilities. For example, there is a variant
of the DFR model where the commutators of the Lorentz covariant coordinates
are not central \cite{Doplicher:1996bb,bdfp_mod}. 

The model proposed here allows for discussing general features of coordinate quantisation. It should be stressed however 
that no direct physical motivations has been devised for this 
particular choice of commutation relations. Although the lack of covariance
has been cured by covariantisation, 
one of our main criticisms to the \(\kappa\)-Minkowski 
model still survives: it is possible to find states (localised around 
any event)  such that all uncertainties
\(\Delta X^\mu\) are simultaneously small at wish. Hence an arbitrarily high
energy density can be transferred in principle to the geometric background
as an effect of localisation alone, which could trap the event in a horizon.
In other words, the model is not stable under localisation alone.

\appendix
\section{}

We recall from \cite{Dabrowski:2010yk} the universal regular
representation of the usual \(\kappa\)-Minkowski relations
\begin{equation}\label{eq:old_kappa}
[X^0_{(0)},X^j_{(0)}]=iX_{(0)}^j,\quad [X^j_{(0)},X^k_{(0)}]=0.
\end{equation}

The regular (Weyl) form of the above
relations is
\begin{equation}
\label{eq:old_kappa_weyl}
e^{ih_\mu X^\mu_{(0)}}e^{ik_\mu X^\mu_{(0)}}=e^{i\phi_\mu(h,k)X^\mu_{(0)}},
\end{equation}
where 
\begin{subequations}
\label{eq:phi}
\begin{align}
\phi^0(h,k)&=h^0+k^0,\\
\vec\phi(h,k)&=
-w(h^0+k^0,h^0)e^{ik_0}\bh
-w(h^0+k^0,k^0)\bk,
\end{align}
\end{subequations}
and \(w(s,t)=\frac{s(e^t-1)}{t(e^s-1)}\) 
\cite{Kosinski:1999dw,Dabrowski:2010yk}. 
Above, \(v,h,k\) are 4-vectors, and
we use the decomposition \(v=(v^0,\bv)\).

Consider the measure 
\[
d\mu(\vec c,s)=\big(\delta(|\vec c|-1)+\delta(|\vec c|)\big)d\vec c\;ds
\]
on \(\mathbb R^3\times\mathbb R\), where \(d\vec c,ds\) are the Lebesgue
measures on \(\mathbb R^3,\mathbb R\), respectively; 
its support is \((S^2\cup\{0\})\times\mathbb R\). On the space
\(L^2(\mathbb R^3\times\mathbb R,d\mu(\vec c,s))\), the operators
\begin{subequations}
\begin{align}
&(X_{(0)}^0\xi)(\vec c,s)=-i\frac{\partial\xi}{\partial s}(\vec c,s),\\
&(X_{(0)}^j\xi)(\vec c,s)=c_je^{-s}\xi(\vec c,s)
\end{align} 
\end{subequations}
fulfil \eqref{eq:old_kappa_weyl}.
Moreover, they contain precisely one representative for every 
equivalence class of irreducible representation of the above relations. 

\bibliographystyle{utphys} 
\bibliography{mybibtex}

\end{document}